\documentclass[11pt,a4paper,final]{article}
\usepackage{a4wide}             
\usepackage{datetime}
\ddmmyyyydate
\usepackage{ifpdf}               
\ifpdf       
\fi
\usepackage{amssymb}            
\usepackage{amsmath}
\usepackage[notref,notcite]{showkeys} 
\ifpdf
\usepackage[pdftex]{graphicx}
\usepackage[pdftex,unicode,implicit]{hyperref}
\hypersetup{%
  pdftitle = {N=2 Super-EYM coloured black holes from defective Lax matrices},
  pdfkeywords = {N=2 Einstein-Yang-Mills, coloured black holes, Lax pair,
    defective matrices}, pdfauthor = {P. Meessen, T. Ort\'{\i}n}, pdfcreator =
  {pdf\LaTeXe\ with package \flqq hyperref\frqq}, pdfproducer = {pdf\LaTeXe\
    with package \flqq hyperref\frqq}, pdfpagemode =
  None, 
  pdffitwindow= true, 
  unicode = true, plainpages = false, colorlinks =
  true, 
  citecolor = black, urlcolor = black, linkcolor = black }

 \else
\usepackage[dvips]{graphicx} \usepackage[unicode,implicit]{hyperref}

\fi
\usepackage{tikz}
\newcommand{\FPAUO}[2]{
\tikz[scale=.13,
      Uniovi/.style={color=green!51!blue, fill=green!51!blue}
 ] {
 \fill[Uniovi] (0,0) circle (10);
 \fill[white] (0,7) circle (1.5);
 \draw[Uniovi] (-2,7.5) rectangle (2,5.5);
 \fill[white] (-0.3,6.6) rectangle (0.3,0);   
 \fill[white] ( -0.9,6.2) rectangle (.9 ,5.6);
 \fill[white] (-1.4, 5.2) rectangle (1.4, 4.6);
 \fill[white] (0,0) ellipse (3.5 and 4);
 \fill[Uniovi] (-2.5,0.3) rectangle (2.5,-0.3);
 \fill[Uniovi] (-2,2.3) rectangle (2,1.7);
 \fill[Uniovi] (-2,-2.3) rectangle (2,-1.7);
 \fill[white] (-4.5,5.5) rectangle (-2.7,4.9);
 \fill[white] (-3.9,6.1) rectangle (-3.3,4.3);
 \fill[white] (4.5,5.5) rectangle (2.7,4.9);
 \fill[white] (3.9,6.1) rectangle (3.3,4.3);
 \foreach \x in { 0,..., 3 }
   \foreach \y in { 0,...,\x}
    {
     \fill[white] (-6-\x*0.7+\y*1.4,3.5-\x *1.97) -- (-5.6-\x*0.7+\y*1.4,2.4-\x *1.97) -- (-6.4-\x*0.7+\y*1.4,2.4-\x *1.97) -- cycle;
     \fill[white] (6-\x*0.7+\y*1.4,3.5-\x *1.97) -- (5.6-\x*0.7+\y*1.4,2.4-\x *1.97) -- (6.4-\x*0.7+\y*1.4,2.4-\x *1.97) -- cycle;
   };
 \draw (0,-6) node[
                               text centered, 
                               color=white, 
                               font={\fontsize{8}{4}\sffamily\selectfont}
                             ] {FPAUO-#1/#2};
}} 
\begin{document}   
~\vspace{-4cm}\begin{flushright}
\small
\FPAUO{14}{09}\\
IFT-UAM/CSIC-15-002\\
\today
\normalsize
\end{flushright}
\begin{center}
{\Large\bf $\mathcal{N}=2$ Super-EYM coloured black holes from defective 
Lax matrices}\\[.5cm]
{\bf Patrick~Meessen$^{\aleph}$}\footnote{E-mail: {\tt meessenpatrick [at] uniovi.es}}
~~{\bf and}~~
{\bf Tom\'as Ort\'{\i}n$^{\beth}$}\footnote{E-mail: {\tt Tomas.Ortin [at] csic.es}}\\[.2cm]
{\em $^{\aleph}$HEP Theory Group, Departamento de F\'{\i}sica, Universidad de Oviedo\\ 
        Avda.~Calvo Sotelo s/n, E-33007 Oviedo, Spain
}\\[.3cm]
{\em $^{\beth}$Instituto de F\'{\i}sica Te\'orica UAM/CSIC, 
        C/ Nicol\'as Cabrera, 13--15\\  C.U.~Cantoblanco, E-28049 Madrid, Spain}
\\[.6cm]
{\small\em Dedicated to the memory of J.-J. Mill\'an Santamar\'{\i}a 
}\\[.6cm]
{\bf abstract}\\
\begin{quote} {\small We construct analytical supersymmetric coloured black
    hole solutions, {\em i.e.\/} non-Abelian black hole solutions that have no
    asymptotic non-Abelian charge but do have non-Abelian charges on the
    horizon that contribute to the Bekenstein-Hawking entropy, to two
    $\mathrm{SU}(3)$-gauged $\mathcal{N}=2$ $d=$ supergravities. The
    analytical construction is made possible due to the fact that the main
    ingredient is the Bogomol'nyi equation, which under the assumption of
    spherical symmetry admits a Lax pair formulation. The Lax matrix needed
    for the coloured black holes must be defective which, even though it is
    the non-generic and less studied case, is a minor hindrance.}
\end{quote}
\end{center}
\vspace{.5cm}

In refs.~\cite{Huebscher:2007hj,Hubscher:2008yz} the structure of the
supersymmetric solutions to $\mathcal{N}=2,d=4$ supergravity coupled to vector
multiplets where a subset of the isometries of the scalar manifold is gauged
\cite{deWit:1984px,Andrianopoli:1996cm}, a theory we shall refer to as
$\mathcal{N}=2,d=4$ Super-Einstein-Yang-Mills, was uncovered. As far as said
characterisation is concerned, the resulting structure is {\em grosso modo}
the same as in the ungauged case, see {\em e.g.\/}
refs.~\cite{Behrndt:1997ny,Meessen:2006tu}, but for two details: the seed
functions $\mathcal{I}^{\Lambda},\mathcal{I}_{\Lambda}$ are in general not
harmonic functions on $\mathbb{R}^{3}$. More to the point, some of the seed
functions (the \textit{magnetic} ones $\mathcal{I}^{\Lambda}$ that have indices
$\Lambda$ in the non-Abelian gauge group and are denoted by $\Phi$) must
satisfy the non-Abelian\footnote{In the ungauged case (and in the ungauged
  directions) the seed functions also have to obey the Bogomol'nyi equation,
  but in its Abelian version. A solution (gauge field) is guaranteed to exist
  for any choice of harmonic seed functions and, for this reason, the presence
  of the Abelian Bogomol'nyi equation is seldom mentioned in the
  literature. In the non-Abelian case, however, one cannot simply give a set
  of seed functions satifying certain conditions: it is necessary to provide
  the accompanying gauge field to determine completely the solution.}
Bogomol'nyi equation on $\mathbb{R}^{3}$, {\em i.e.\/}
\begin{equation}
  \label{eq:1}
  \star\mathsf{F}\; =\; \mathsf{D}\Phi \; ,
\end{equation}
whereas the other half (the \textit{electric} ones $\mathcal{I}_{\Lambda}$)
must satisfy an involved equation whose explicit form depends on the knowledge
of the solutions to eq.~(\ref{eq:1}).
\par
Some supersymmetric, analytical spherically symmetric solutions to various
models were constructed and analysed in
refs.~\cite{Huebscher:2007hj,Meessen:2008kb,Hubscher:2008yz} and, recently,
some multi-object solutions were presented in ref.~\cite{Bueno:2014mea}.
Apart from the globally regular solutions describing 't Hooft-Polyakov
monopoles in supergravity, the most elusive and intriguing solutions in those
references describe an $\mathrm{SU}(2)$ black hole whose non-Abelian gauge
fields have no asymptotic colour charge and the physical scalars have no
asymptotic v.e.v. The near-horizon solution, however, is completely specified
in terms of the colour charge of the gauge fields on the horizon, which does
not vanish and contributes to the Bekenstein-Hawking entropy.  This behaviour
is reminiscent of the Bartnik-McKinnon particle solution to $\mathrm{SU}(2)$
Einstein-Yang-Mills (EYM) theory \cite{Bartnik:1988am}, which is a discrete
family of spherically symmetric, globally regular solutions with a YM
connection that has no asymptotic colour charge. The Bartnik-McKinnon
solutions were generalised to black-hole solutions in
refs.~\cite{Bizon:1990sr,Kuenzle:1990is,Volkov:1990sva} and they are given
the, perhaps counter-intuitive, name of \textit{coloured black holes} and,
accordingly, the elusive solution mentioned above is called an $\mathcal{N}=2$
\textit{Super-EYM (SEYM) coloured black hole}.
\par
Observe that the Bartnik-McKinnon particle and their generalisations are only
known numerically, and analytical examples are hard to come by. As far as we
know, 
apart from the sugra coloured black hole of
refs.~\cite{Meessen:2008kb,Hubscher:2008yz}, the only other analytical
coloured black hole was recently constructed by Fan {\&} L\"u
\cite{Fan:2014ixa} and is a solution to conformal gravity coupled to
$\mathrm{SU}(2)$ YM. This last coloured black hole, like the coloured black
holes of EYM, is ``pure'' in the sense that for their existence no other gauge
field needs to be turned on. This is in contradistinction to the coloured
black hole in supergravity where the regularity of the event horizon demands
an additional Abelian gauge field to be active. Even so, it is, within the
class of theories and (supersymmetric) solutions considered, the closest one
can get to a coloured black hole.
\par
Seeing the interesting properties of the coloured black holes, however, it
would be interesting to have more analytical examples; in the
$\mathcal{N}=2,d=4$ case we are quite fortunate as the main ingredient of
their construction is the Bogomol'nyi equation and as was shown by Leznov {\&}
Saveliev in ref.~\cite{Leznov:1980tz}, the $\mathrm{SU}(N)$ Bogomol'nyi
equation, under the assumption of (so-called maximal) spherical symmetry, is
an integrable system related to the $\mathrm{SU}(N)$ Toda molecule.  The
general solution of the $\mathrm{SU}(N)$ Toda molecule is known
\cite{Koikawa:1981xg,Farwell:1982du,Farwell:1983sx,Anderson:1995sz} and can be
derived from a Lax pair representation for the relevant equations, with a Lax
matrix that is assumed to be diagonalisable. As we will see, the solutions to
the spherically symmetric Bogomol'nyi equations that are needed to construct
coloured black holes are associated to a Lax matrix that is defective or, said
differently, manifestly non-diagonalisable.  The ones we are after can be
obtained from the generic solutions as limiting cases (see {\em e.g.\/}
ref.~\cite{Lu:1996jr}) or by the integration algorithms presented in
refs.~\cite{Fre:2009dg,Chemissany:2009af} and \cite{Chemissany:2010zp}. Here
we will adapt Koikawa's derivation of the solution \cite{Koikawa:1981xg} to
the case of a defective Lax matrix: this is done by relating the Lax pair to
the standard Lax evolution equation on the generalised eigenvectors of the
defective Lax matrix. The generalised eigenvectors are the vectors with
respect to~which the Lax matrix takes on the Jordan block form and are natural
objects to use when the Lax matrix is defective.  For the small $N$
$\mathrm{SU}(N)$ solutions we are going to construct ($\mathrm{SU}(2)$ and
$\mathrm{SU}(3)$) this gives a convenient and easy to understand way of
constructing the solutions, clarifying the general procedure of
refs.~\cite{Fre:2009dg,Chemissany:2009af} and \cite{Chemissany:2010zp}.
\par
Once we have constructed the relevant solutions of the $\mathrm{SU}(N)$
Bogomol'nyi equations, we will use them to construct {\em bona fide} coloured
black hole solutions to $\mathcal{N}=2,d=4$ SEYM theories. In this case we
will limit ourselves to the construction of coloured black holes in the
$\overline{\mathbb{CP}}^{8}$ and the $\mathbb{C}$-magic models, both of which
allow for an $\mathrm{SU}(3)$ gauging.  In both cases, we will see that, as in
the $\mathrm{SU}(2)$ coloured black holes presented in
refs.~\cite{Meessen:2008kb,Hubscher:2008yz,Bueno:2014mea}, there is no major
obstruction in their construction. In fact, it should be more or less clear
that coloured black-hole solutions abound in $\mathcal{N}=2,d=4$ SEYM
theories.
\par
The outline of this article is as follows: in sec.~\ref{sec:SpherBogo} we
shall discuss the form of the gauge connections and Higgs fields compatible
with spherical symmetry, relate the $\mathrm{SU}(N)$ Bogomol'nyi equation
under the assumption of spherical symmetry to the $\mathrm{SU}(N)$ Toda
molecule and give the Lax pair. We will then discuss the appropriate boundary
conditions for the coloured solutions we are interested in and argue that they
correspond to a defective Lax matrix, {\em i.e.\/} to a non-diagonalisable
matrix. In sec.~\ref{sec:defective} we will discuss the general mechanism
needed for extracting the solutions and apply it to the cases of
$\mathrm{SU}(2)$ and $\mathrm{SU}(3)$. In sec.~\ref{sec:CMagic} we will then
embed the $\mathrm{SU}(3)$ solution in the $\mathrm{SU}(3)$-gauged
$\overline{\mathbb{CP}}^{8}$ and $\mathbb{C}$-magic models and analyse the
gravitational solution.  Section \ref{sec:Concl} contains some conclusions and
comments.
\section{Spherically symmetric solutions to the Bogomol'nyi equations}
\label{sec:SpherBogo}
The derivation is best done using Hermitean generators, which means that we
use the definitions
\begin{equation}
  \label{eq:WB1}
  \mathsf{D}\Phi \; =\; d\Phi -i\left[ \mathsf{A},\Phi\right] \;\; ,\;\;
  \mathsf{F} \; =\; d\mathsf{A} \ -\ i\ \mathsf{A}\wedge \mathsf{A} \; ,
\end{equation}
where $\mathsf{A}$ and $\Phi$ are in $\mathfrak{su}(n+1)$'s fundamental
representation; for convenience we have taken the coupling constant to be one.
\par
In the radial gauge, {\em i.e.\/} in the gauge where the radial component of
the gauge connection vanishes, the form of the fields compatible with maximal
spherical symmetry are given by (see {\em e.g.\/}
refs.~\cite{zbMATH03141695,Wilkinson:1978zh,Ganoulis:1981sx,Kuenzle:1991wa})
\begin{eqnarray}
  \label{eq:WB2a}
  \Phi 
& = & 
\tfrac{1}{2}\, \mathrm{diag}
\left(
\phi_{1}(r)\ ,\ \phi_{2}(r)-\phi_{1}(r)\ ,\ \ldots\ ,\  \phi_{n}(r)-\phi_{n-1}(r)\ ,\ -\phi_{n}(r)
\right) \; , 
\\
& & \nonumber\\
  \label{eq:WB2b}
  \mathsf{A}  
& = & 
J_{3}\ \cos (\theta )d\varphi 
     \ +\ \tfrac{i}{2}\left[ C -C^{\dagger}\right]\ d\theta
     \ +\ \tfrac{1}{2}\left[ C+C^{\dagger}\right]\ \sin (\theta )d\varphi \; ,
\end{eqnarray}
where
\begin{equation}
  \label{eq:32}
  J_{3}
\; =\; 
\tfrac{1}{2}\mathrm{diag}( n, n-2, \ldots , 2-n, -n) \; ,
\end{equation}
corresponds to a spin $n/2$ irrep of $\mathfrak{su}(2)$ and is the maximal
embedding of $\mathfrak{su}(2)$ into $\mathfrak{su}(n+1)$; $C$ is the real
upper-triangular matrix
\begin{equation}
  \label{eq:WB3}
  C \ =\ \left(\begin{array}{ccccc}
             0 & a_{1}(r) & 0 & \cdots & 0 \\
             0 & 0 & a_{2}(r) & \cdots & 0\\
             \vdots & \vdots & \vdots & \vdots & \vdots \\
             0 & 0 & \cdots & a_{n-1}(r) & 0 \\
             0 & 0 & \cdots & 0 & a_{n}(r) \\
             0 & 0 & \cdots & 0 & 0 
         \end{array}\right)\; .
\end{equation}
A small calculation using the fact that $\left[ J_{3},C\right] = C$, then
shows that the Bogomol'nyi equation (\ref{eq:1}) reduces to the following
system
\begin{eqnarray}
  \label{eq:51}
  \partial_{r}C 
& = & 
\left[\ \Phi\ ,\ C\ \right] \; , 
\\
& & \nonumber\\
 \label{eq:51b}
2r^{2}\partial_{r}\Phi 
& = & 
\left[\ C\ ,\ C^{\dagger}\ \right]\; -2J_{3}\; . 
\end{eqnarray}
\par
The starting point of Koikawa's construction are eqs.~(\ref{eq:51}) and
(\ref{eq:51b}). First, one redefines\footnote{$F$ should not be confused with
  the gauge field strength $\mathsf{F}.$}
\begin{equation}
  \label{eq:52}
  \Phi \; =\; \Psi \ +\ J_{3}/r \;\;\; ,\;\;\;
  C \; =\; rF \; ,
\end{equation}
turning eqs.~(\ref{eq:51}) and (\ref{eq:51b}) into
\begin{eqnarray}
\label{eq:53}
\dot{F} 
& = & 
\left[\ \Psi\ ,\ F\ \right] \; , 
\\
& & \nonumber\\
\label{eq:53b}
2\dot{\Psi} 
& = & 
\left[\ F\ ,\ F^{T}\ \right] \; , 
\end{eqnarray}
where clearly $\Psi^{T}=\Psi =\Psi^{\dagger}$ and an overdot means derivative
with respect to~$r$.
\par
After this redefinition, the link to the Toda molecule can be easily
established \cite{Farwell:1983sx}: in terms of the components $\psi_{i}$ of
$\Psi$ and $f_{i}$ of $F$, the above equations read
\begin{equation}
  \label{eq:112}
  \dot{\psi}_{i}\; =\; f_{i}^{2} \;\;\; ,\;\;\;
  2\dot{f}_{i}\; =\; f_{i}\ A_{ij}\psi_{j} \; ,
\end{equation}
where $A$ is $\mathfrak{su}(n+1)$'s Cartan matrix.  Defining the new variables
$T_{i}(r)$ by 
\begin{equation}
  \label{eq:113}
  f_{i}\; =\; \exp\left( -\tfrac{1}{2}\ A_{ij} T_{j}\right) \; ,
\end{equation}
we see from the $f$-equations that $\psi_{i}=-\dot{T}_{i}$. 
Upon substitution into the $\psi$-equation we find that the $T_{i}$ variables
satisfy the equations
\begin{equation}
  \label{eq:114}
  \ddot{T}_{i} \; =\; -\exp\left( -A_{ij}T_{j}\right) \; ,
\end{equation}
which are known as the equations of motion of the $\mathrm{SU}(n+1)$ Toda
molecule.
\par
The second step in Koikawa's construction \cite{Koikawa:1981xg} is the
definition of the new objects $\mathsf{L},\mathsf{M}$ from $\Psi$ and $F$,
which form a Lax pair, \textit{i.e.}
\begin{equation}
  \label{eq:54}
  \left.
    \begin{array}{lcl}
        \mathsf{L} & \equiv& \Psi \ +\ \tfrac{i}{2}\left( F+F^{T}\right) \\
        & & \\
        \mathsf{M}& \equiv& \tfrac{i}{2}\left( F^{T}-F\right)
     \end{array}
  \right\} \hspace{.5cm}\longrightarrow\hspace{.5cm}
  \dot{\mathsf{L}}\; =\; \left[\ \mathsf{M}\ ,\ \mathsf{L}\ \right] \; ,
\end{equation}
The existence of a Lax pair immediately implies that the quantities
(``charges'') $\mathcal{C}_{(k)}\equiv\mathrm{Tr}\left(\mathsf{L}^{k}\right)$
are constants of motion.  Observe that even though $\mathsf{L}$ is symmetric
and complex, whereas $\mathsf{M}$ is purely imaginary and anti-symmetric, the
conserved charges $\mathcal{C}_{(k)}$ are in fact real functions because
$\mathsf{L}$ and $\mathsf{M}$ satisfy the relations
\cite{Ganoulis:1981sx}
\begin{equation}
  \label{eq:17}
  \mathsf{L}^{*}\; =\; e^{i\pi J_{3}}\ \mathsf{L}\ e^{-i\pi J_{3}} 
  \;\;\; ,\;\;\;
  \mathsf{M}^{*}\; =\; e^{i\pi J_{3}}\ \mathsf{M}\ e^{-i\pi J_{3}} \; .
\end{equation}
\par
We  are interested in Higgs fields that behave asymptotically as
\begin{equation}
\label{eq:116}
\Phi 
\; \sim\; 
\Phi_{\infty}\ +\ \frac{\mathrm{P}}{r} \ +\ \mathcal{O}(r^{-2})\; ,
\end{equation}
where $\mathrm{P}$ is the colour charge and one can take $\left[
  \Phi_{\infty},\mathrm{P}\right] =0$.\footnote{We can always perform a
  constant gauge transformation to diagonalise $\Phi_{\infty}$ and an
  $r$-dependent one to diagonalise $\mathrm{P}$.} According to
eq.~(\ref{eq:53b}), in order for the above asymptotic behaviour to be possible
we must have\footnote{ In components this reads $f_{i}\sim s_{i}/r$ for some
  constants $s_{i}$.}
\begin{equation}
\label{eq:2}
F
\; \sim \; 
\frac{\mathrm{S}}{r}
\hspace{1cm}\mbox{with}\;\;\;
\left[\mathrm{S},\mathrm{S}^{T}\right] 
\; =\; 
2\left(\ J_{3}\ -\ \mathrm{P}\ \right) \; .
\end{equation}
Eq.~(\ref{eq:53}) then implies that 
\begin{equation}
  \label{eq:3}
  \left[\ \Phi_{\infty}\ ,\ \mathrm{S}\ \right] \; =\; 0 
  \hspace{.5cm}\mbox{and}\hspace{.5cm}
  \left[\ \mathrm{P}\ ,\ \mathrm{S}\ \right] \; =\; 0 \; .
\end{equation}
Defining then a coloured solution to the Bogomol'nui equation as one for which the colour charge is zero, {\em i.e.\/} $\mathrm{P}=0$, we see from 
eq.~(\ref{eq:2}) that $s_{i}^{2}= i(n+1-i)$, so that $\mathrm{S}$ has no vanishing entries; this immediately implies by virtue of
eq.~(\ref{eq:3}) that $\Phi_{\infty}=0$. The conclusion is that if, in the maximally spherically symmetric Ansatz, we want to 
describe coloured solutions, defined as solutions having $\mathrm{P}=0$, then we must look at the class of solutions with $\Phi_{\infty}=0$.
Observe, however, that it is possible to have solutions with $\Phi_{\infty}=0$ but $\mathrm{P}\neq 0$.
\par
{}For the conserved charges $\mathcal{C}_{(k)}$ we see that
\begin{equation}
  \label{eq:115}
  \mathcal{C}_{(k)} 
     \; =\; \lim_{r\rightarrow\infty}\mathrm{Tr}\left(\mathsf{L}^{k}\right)
     \; =\; \mathrm{Tr}\left( \Phi_{\infty}^{k}\right)
     \;\;\;\xrightarrow{\;\;\mbox{coloured solutions}\;\;}\;\;\;
     \mathcal{C}_{(k)} \; =\; 0 \; .
\end{equation}
Even though $\mathsf{L}$ is a complex symmetric matrix it allows for eigenvalues and if we take into account the fact that 
the characteristic polynomial $\det (\mathsf{L}-\lambda\ \mathrm{Id})$ can be expanded as a polynomial in $\lambda$ and the $\mathcal{C}_{(k)}$, 
we reach the conclusion that a coloured solution is such that all eigenvalues are zero: this is only possible for non-trivial solutions if
$\mathsf{L}$ is not diagonalisable, whence an $\mathsf{L}$ corresponding to a
coloured solution must be a defective matrix, \textit{i.e.}~one that can be
brought by an $r$-dependent similarity transformation to a Jordan block form
with only zero eigenvalues.
\par
Observe that for our purposes we also might want to impose the condition that
around $r=0$\footnote{This is the location of the would-be event horizon in
  the full gravitational solutions built from these solutions of the
  Bogomol'nyi equation.} we want the solution to be Coulombic, meaning that
\begin{equation}
\label{eq:118}
\lim_{r\rightarrow 0}\Phi 
\; \sim\; 
\frac{\mathrm{Q}}{r}\ +\ \Phi_{0} \ +\ \mathcal{O}(r)\; , 
\end{equation}
as any higher order singularity would (probably) be uncompensable in an
$\mathcal{N}=2$ supergravity setting. The solutions that we have constructed,
however, automatically have this behaviour.
\section{Coloured solutions from defective Lax matrices}
\label{sec:defective}
The Lax equation (\ref{eq:54}) is the integrability condition of the so-called
Lax equations\footnote{ The vector $\vec{v}$ is $r$-dependent, but we refrain
  from writing $\vec{v}(r)$ for the ease of reading.}
\par
\begin{eqnarray}
\label{eq:62}
\mathsf{L}\vec{v} 
& = & 
\lambda\ \vec{v} \; , 
\\
& & \nonumber \\
\label{eq:62b}
\dot{\vec{v}}
& = & 
\mathsf{M}\ \vec{v} \; ,
\end{eqnarray}
where $\lambda$ is $r$-independent. As was said before, the matrix
$\mathsf{L}$ is symmetric but complex and as such it need not be
diagonalisable. But since this possibility is not excluded either, let us
suppose first that all the eigenvalues\footnote{As we are dealing with the
  $\mathrm{SU}(n+1)$ Bogomol'nyi equation, the sum of all the eigenvalues must
  be zero.} of $\mathsf{L}$ are real and different \cite{Koikawa:1981xg}.
This means that there are $n$ eigenvalues $\lambda_{i}$ ($i=1,\ldots ,n$) and
$n$ corresponding eigenvectors $\vec{v}_{i}$, which due to eq.~(\ref{eq:62b})
can be taken to be orthonormal, $\vec{v}_{i}\cdot\vec{v}_{j}=\delta_{ij}$, and
complete $\mathrm{Id}\ =\ \sum_{i}\ \vec{v}_{i}\vec{v}_{i}^{\, T}$.  We can then
immediately write down the spectral decomposition of $\mathsf{L}$
\begin{equation}
\label{eq:63}
\mathsf{L}
\; =\; 
\sum_{i}\ \lambda_{i}\ \vec{v}_{i} \vec{v}_{i}^{\, T} \; ,
\end{equation}
and express the resolvent of $\mathsf{L}$ as
\begin{equation}
\label{eq:64}
\mathsf{R}(\mu ) 
\; \equiv\; 
\left(\mathsf{L}-\mu\ \mathrm{Id}\right)^{-1}
\; =\; 
\sum_{i}\ \frac{\vec{v}_{i}\vec{v}_{i}^{\, T}}{\lambda_{i}-\mu} \; .
\end{equation}
\par
The Lax equations can be turned into $n$ first order differential equations by
projection. Let us consider the $\mathrm{SU}(2)$ case with a basis
$\{\vec{e}_{1},\vec{e}_{2}\}$, define $x_{i}\equiv
\vec{e}_{1}\cdot\vec{v}_{i}$ and use the definitions of $\mathsf{L}$ and
$\mathsf{M}$ eqs.~(\ref{eq:54}) to see that
$\mathsf{M}_{12}=-\mathsf{L}_{12}$. Whence, from eq.~(\ref{eq:62b}), we have
that $\dot{x}_{i}\ =\ \mathsf{M}_{12}\ \vec{e}_{2}\cdot\vec{v}_{i}
=-\mathsf{L}_{12}\ \vec{e}_{2}\cdot\vec{v}_{i}$. Using these expressions in
the projection of eq.~(\ref{eq:62}) onto $\vec{e}_{1}$ we find
\begin{equation}
  \label{eq:14}
\dot{x}_{i}
\; =\; 
\mathsf{L}_{11}x_{i}\ -\ \lambda_{i}\ x_{i} 
\hspace{1cm}\mbox{with}\;\; 
\mathsf{L}_{11}
\ =\ 
\sum_{j}\ \lambda_{j}x_{j}^{2}\; .
\end{equation}
By substituting $x_{i}=e^{-\lambda_{i}r}\ a_{i}\ u(r)$, where $a_{i}$ is some
constant, we see that the general solution to eqs.~(\ref{eq:14}) is given by
\begin{equation}
  \label{eq:15}
  \frac{1}{u^{2}}\; =\; \aleph\ +\ \sum_{j}\ a_{j}^{2}\ e^{-2\lambda_{j} r}
  \hspace{.4cm}\xrightarrow{\;\;\mbox{completeness}\;\;}\hspace{.4cm}
  \frac{1}{u^{2}}\; =\; \sum_{j}\ a_{j}^{2}\ e^{-2\lambda_{j} r} \; ,
\end{equation}
where the completeness alluded to in the equation follows from the projection
onto the $11$-direction of the completeness relation of the $\vec{v}_{i}$.  Up
to this point we can, by using eqs.~(\ref{eq:32}) and (\ref{eq:54}), deduce
that
\begin{equation}
\label{eq:16}
\psi_{1}
\; =\; 
2\mathsf{L}_{11}
\; =\; 
2\frac{\sum_{i}\ \lambda_{i}a_{i}^{2}\ e^{-2\lambda_{i}r}}{\sum_{j}\ 
a_{j}^{2}\ e^{-2\lambda_{j}r}} \; .
\end{equation}
\par
Even though it may seem that in order to construct the full solution one would
have to find all the components of the eigenvectors, this is not really
needed: we can calculate the $11$-component of the resolvent,
$\mathsf{R}_{11}(\mu )$, in two different ways, namely by projection of the
spectral representation in eq.~(\ref{eq:64}) and also by using Cramer's rule
to invert the matrix $\mathsf{L}-\mu\mathrm{Id}$, given eq.~(\ref{eq:52}). The
comparison of these two expressions is enough to fix all the components. We
refer the interested reader to Koikawa's article \cite{Koikawa:1981xg} for the
final result. As the reader will see, using the resolvent to fix the complete
solution is especially easy for the coloured solutions, at least for the
low-$n$ cases we are interested in.
\par
In the case that the eigenvalues are degenerate we must distinguish between
two cases: the non-defective case and the defective case.  In the
non-defective case in which $\mathsf{L}$ can still be diagonalised even though
the eigenvalues are degenerate, a complete set of eigenvectors does still
exist and the above method can be applied.
\par
In the defective case in which the matrix can be brought to a Jordan block
form but not a diagonal form, we need to introduce the so-called {\em
  generalised eigenvectors}: suppose we are given an eigenvalue
$\lambda_{\star}$ which corresponds to some $k\times k$ Jordan
block;\footnote{The generalisation to more Jordan blocks, even with the same
  eigenvalue, is straightforward, but we refrain from describing it here in order
  not to clutter the equations with too many indices.} in this case there is
only one eigenvector $\vec{v}_{\star}$ such that
\begin{equation}
\mathsf{L}\vec{v}_{\star}=\lambda_{\star}\vec{v}_{\star}\; .
\end{equation}
We can, however, construct a basis for the complete eigenspace with
\textit{generalised eigenvectors} $\vec{w}(m)$ where $m=1,\ldots ,k$ and
$\vec{w}(k)\equiv \vec{v}_{\star}$ defined by the property
\begin{equation}
\label{eq:65}
\vec{w}(m+1)
\; =\; 
\left( \mathsf{L}-\lambda_{\star}\mathrm{Id}\right)\ \vec{w}(m) 
\; \equiv\; 
\mathsf{N}\vec{w}(m) \; ,
\end{equation}
with the understanding that $\vec{w}(k+1)=0$.  This equation makes sense as
the matrix $\mathsf{N}$ restricted to the generalised eigenspace corresponding
to the eigenvalue $\lambda_{\star}$ is nil-potent of degree $k$, {\em i.e.\/}
\begin{equation}
\mathsf{N}^{k}=0\;.
\end{equation}
\par
The integrability condition of  the evolution equation
\begin{equation}
\label{eq:66}
\dot{\vec{w}}(m) \; =\; \mathsf{M}\vec{w}(m) \; ,
\end{equation}
with the definitions (\ref{eq:65}) is nothing more than the Lax pair equation
(\ref{eq:54}). Said differently: if we find a solution to the system formed by
eqs.~(\ref{eq:65}) and (\ref{eq:66}), we automatically obtain a solution to
the original equations (\ref{eq:51}) and (\ref{eq:51b}).
\par
The basis $\{\vec{\omega}(m)\}$ is by construction complete but the details
vary with respect to~the non-degenerate case as, for example,
$\vec{w}(k)\cdot\vec{w}(k)=0$,\footnote{The $\vec{w}(m)$ are complex.} so that
the relevant spectral representations will not be as simple as in
eq.~(\ref{eq:63}) and (\ref{eq:64}). First of all, since $\mathsf{M}$ is
antisymmetric the inner products of the generalised eigenvectors are
$r$-independent, {\em i.e.\/}
\begin{equation}
\label{eq:67}
\partial_{r}\left[ \vec{w}(m)\cdot\vec{w}(n)\right] 
\; =\; 
0\, ,
\hspace{.4cm}\mbox{whence}\hspace{.4cm}
\vec{w}(m)\cdot\vec{w}(n) 
\; =\; 
\left.\vec{w}(m)\cdot\vec{w}(n)\right|_{r=0} \; .
\end{equation}
Second, one finds that $\vec{w}(m)\cdot\vec{w}(n)\ =\
\vec{w}(m-1)\cdot\vec{w}(n+1)$ so this product only depends on the sum of the
indices $m+n$. This, together with $\mathsf{N}\vec{w}(k)=0$, implies that
\begin{equation}
  \label{eq:68}
  \vec{w}(m)\cdot\vec{w}(n) \; =\; 0 \;\;\;\;\; \mbox{if $m+n> k+1$.}
\end{equation}
In fact, since eq.~(\ref{eq:65}) determines $\vec{w}(m)$ ($m\neq k$) up to
terms proportional to $\vec{w}(k)$ we can arrange things such that
\begin{equation}
  \label{eq:69}
  \vec{w}(m)\cdot\vec{w}(n) \; =\; \left\{
       \begin{array}{ccc}
            0 &\;\;\; :\;\;\;&  m+n\ \neq\ k+1\; , \\
            1 & :& m+n\ =\ k+1\; .
       \end{array}
  \right.
\end{equation}
Having deduced the relevant inner products of the generalised eigenvectors one
can write down the completeness relation on this eigenspace as
\begin{equation}
\label{eq:70}
\left.\mathrm{Id}\right|_{\lambda_{\star}} 
\; =\; 
\sum_{m=1}^{k}\ \vec{w}(k+1-m)\vec{w}(m)^{T} \; ,
\end{equation}
and the (generalised) spectral representation of $\mathsf{L}$ reads
\begin{equation}
\label{eq:71}
\left.\mathsf{L}\right|_{\lambda_{\star}} 
\; =\; 
\lambda_{\star}\ \sum_{m=1}^{k}\ \vec{w}(k+1-m)\vec{w}(m)^{T} 
                   \ +\ \sum_{m=2}^{k}\ \vec{w}(k+2-m)\vec{w}(m)^{T} \; .
\end{equation}
Finally, the restriction to this eigenspace of the resolvent $\mathsf{R}(\mu)$
is
\begin{equation}
\label{eq:72}
\left.\mathsf{R}(\mu )\right|_{\lambda_{\star}} 
\; =\; 
\sum_{n=1}^{k} \frac{(-1)^{n+1}}{(\lambda_{\star}-\mu)^{n}} 
\sum_{m=n}^{k}\  \vec{w}(k+n-m)\vec{w}(m)^{T} \; ,
\end{equation}
\par
Having set up the relevant algebraic structures, the deduction of the
solutions follows the same route as outlined above.
\subsection{$\mathrm{SU}(2)$'s coloured solution revisited}
\label{sec:DefSU2}
Given that the sum of all eigenvalues must be zero, there are two cases to be
considered, namely the non-degenerate case
$(\lambda_{1},\lambda_{2})=(-\lambda ,\lambda )$ and
$(\lambda_{1},\lambda_{2})=(0,0)$.  The former was treated in Koikawa's
article and we will focus on the latter, which must correspond to a defective
set-up with $\lambda_{\star}=0$ as otherwise we would be dealing with a
trivial gauge field.
\par
The only independent non-trivial conserved charge reads
\begin{equation}
\label{eq:5}
\mathcal{C}_{(2)} 
\; =\; 
\tfrac{1}{2}\left( \psi_{1}^{2}\ -\ f_{1}^{2}\right) \; ,
\end{equation}
which is more than enough to deduce that coloured solutions are such that
$\psi_{1}=\pm f_{1}$, from which the solution can be constructed immediately
using eqs.~(\ref{eq:112}). For illustrative purposes, however, we will outline
Koikawa's construction anyway.
\par
Since we are in the $k=2$ case we have that $\mathsf{L}_{12}= \tfrac{i}{2}f_{1} =
-\mathsf{M}_{12}$. Then, the projection of eq.~(\ref{eq:66}) along
$\vec{e}_{1}$ can be rewritten as
\begin{equation}
\label{eq:73}
\dot{w}_{1}(m) 
\; =\; 
-\mathsf{L}_{12}w_{2}(m) \; . 
\end{equation}
The construction equation (\ref{eq:65}) gives, for the r.h.s.~of the above
equations
\begin{equation}
  \begin{array}{rcl}
\mathsf{L}_{12}w_{2}(1)  
& = &
-L_{11} \omega_{1}(1) +\omega_{1}(2)\; ,
\\
& & \\
\mathsf{L}_{12}w_{2}(2)  
& = &
-L_{11} \omega_{1}(2)\; ,
\end{array}
\end{equation}
and eq.~(\ref{eq:71}) gives
\begin{equation}
\mathsf{L}_{11} = (\omega_{1}(2))^{2}\, .
\end{equation}
Then, introducing the abbreviations $x\equiv \omega_{1}(2)$ and
$y=\omega_{1}(1)$, eqs.~(\ref{eq:73}) take the form
\begin{eqnarray}
  \label{eq:74}
  \dot{x} & =& x^{3} \; , \\
  \label{eq:74b}
  \dot{y} & =& x^{2}y\ -\ x \; ,
\end{eqnarray}
whose solution is
\begin{equation}
  \label{eq:75}
  x^{-2}\; =\; -2(r+b) \hspace{.4cm},\hspace{.4cm}
  y \; =\; -(r+a)\ x(r) \; ,
\end{equation}
where $a$ and $b$ are integration constants.
\par
In this simple case the resolvent can be straightforwardly calculated to be
\begin{equation}
  \label{eq:90}
  \mathsf{R}(\mu )\; =\; 
  \frac{-1}{2\mu^{2}-\mathcal{C}_{(2)}}
  \left(
     \begin{array}{cc}
               2\mu + \psi_{1} \; &\; if_{1} \\
               if_{1} \; &\; 2\mu -\psi_{1} 
     \end{array}
  \right) \; ,
\end{equation}
and for a coloured solution we must have
\begin{equation}
\label{eq:91}
\mathsf{R}_{11}(\mu ) 
\; =\; 
-\frac{1}{\mu}\; -\ \frac{\psi_{1}}{2\mu^{2}} \; . 
\end{equation}
On the other hand, from the spectral representation of the resolvent in
eq.~(\ref{eq:72}) we get
\begin{equation}
\label{eq:92}
\mathsf{R}_{11}(\mu) 
\; =\; 
-\frac{2xy}{\mu}\; -\; \frac{x^{2}}{\mu^{2}}\; .
\end{equation}
Comparing both expressions we first of all see that
\begin{equation}
  \label{eq:93}
  1\; =\; 2xy
    \; =\; \frac{r+a}{r+b} \; , 
\end{equation}
whence $b=a$.  Secondly we see that $\psi_{1} = 2x^{2} = -(r+a)^{-1}$, which
can be used to deduce $f_{1}$ from the constraint $\mathcal{C}_{(2)}=0$.
\par
The final ingredient in the construction is the imposition of the relevant
regularity conditions. In this case we just have to require the Higgs field to
be regular on the interval $r\in (0,\infty )$, which implies that $a\geq
0$. In order to make contact with the solution in
refs.~\cite{Meessen:2008kb,Hubscher:2008yz} we redefine $a =\lambda^{-1}$ and
find
\begin{equation}
  \label{eq:4}
  \psi_{1}\; =\; -\frac{\lambda}{1+\lambda r}\; =\; \pm f_{1} 
  \;\;\;\;\xrightarrow{\;\;\mbox{which by eq.~(\ref{eq:52}) implies}\;\;\;}\;\;\;\;
  \phi_{1}\; =\; \frac{1}{r(1+\lambda r)} \; .
\end{equation}
This solution corresponds to the coloured solution found by Protogenov in
ref.~\cite{Protogenov:1977tq} and which was used in
refs.~\cite{Meessen:2008kb,Hubscher:2008yz,Bueno:2014mea} to construct
coloured black hole solutions to various $\mathcal{N}=2$ EYM theories; the
parameter $\lambda$ is a hair parameter that doesn't show up in the coloured
black hole's asymptotic nor near-horizon behaviour.
\subsection{$\mathrm{SU}(3)$'s coloured solutions}
\label{sec:DefSU3}
In the $\mathrm{SU}(3)$ case there are two coloured possibilities: a pure
$3\times 3$ Jordan block or a $2\times 2$ Jordan block; this last case will be
briefly treated at the end of this section, and will not lead to a coloured
solution.
\par
The explicit forms for the conserved charges are
\begin{eqnarray}
\label{eq:99}
\mathcal{C}_{(2)} 
& = & 
\tfrac{1}{2}
\left(\
 \psi_{1}^{2}\ -\ \psi_{1}\psi_{2}\ +\ \psi_{2}^{2} \ -\ f_{1}^{2}\ -\ f_{2}^{2}
 \ \right)\; , 
\\
& & \nonumber \\
\label{eq:99b}
\mathcal{C}_{(3)} 
& = & 
\tfrac{3}{8}
\left(\
\psi_{1}f_{2}^{2}\ -\ \psi_{2}f_{1}^{2}\ +\ \psi_{2}\psi_{1}^{2}
\ -\ \psi_{1}\psi_{2}^{2}
\ \right) \; ,
\end{eqnarray}
from which, unlike the $\mathrm{SU}(2)$ case, the coloured solutions are not
too easy to deduce.
\par
In the case that $\mathsf{L}$ corresponds to a $3\times 3$ Jordan
block (so $\lambda_{\star}=0$) its spectral representation reads 
\begin{equation}
\label{eq:76}
\mathsf{L} 
\; =\; 
\vec{w}(3)\vec{w}(2)^{T}\ +\ \vec{w}(2)\vec{w}(3)^{T}
\hspace{.4cm}\mbox{whence}\hspace{.4cm}
\tfrac{1}{2}\psi_{1}
\ =\ 
\mathsf{L}_{11} 
\; =\; 
2 xy \; ,
\end{equation}
where we defined $x\equiv \vec{e}_{1}\cdot\vec{w}(3)$, $y\equiv
\vec{e}_{1}\cdot\vec{w}(2)$ and $z\equiv \vec{e}_{1}\cdot\vec{w}(1)$.
\par
Making use of the same technique as in the foregoing section, we find the
system of equations
\begin{eqnarray}
  \label{eq:77}
  \dot{x} & =& 2x^{2}y \; , \\
  \label{eq:77b}
  \dot{y} & =& 2y^{2}x \ -\ x \; , \\
  \label{eq:77c}
  \dot{z} & =& 2xyz \ -\ y \; .
\end{eqnarray}
This system can be solved by the Ansatz $y = f(r)x\, ,\,\,\, z=g(r)x$. Upon
this substitution, the $y$-equation implies that $f(r) = -(r+a)$ and the
$z$-equation then implies that $\dot{g}=-f$ whence
$g(r)=\tfrac{1}{2}[(r+a)^{2}+b]$. The $x$-equation then implies that
\begin{equation}
\label{eq:78}
\frac{1}{x^{2}} 
\; =\; 
2\left[ (r+a)^{2}+c\right] \; ,
\end{equation}
from which we see that
\begin{equation}
  \label{eq:79}
  \psi_{1}\; =\; -2\frac{r+a}{(r+a)^{2}+c} \; .
\end{equation}
\par
In order to find the full solution we compute the $11$-component of the
resolvent in eq.~(\ref{eq:72}), which reads
\begin{equation}
\label{eq:94}
\mathsf{R}_{11}
\; =\; 
-\frac{2xz+y^{2}}{\mu}
\; -\; \frac{2xy}{\mu^{2}}
\; -\; \frac{x^{2}}{\mu^{3}} \; .
\end{equation}
Calculating the same component using Cramer's rule we see that\footnote{
  This can be easily derived using
  \begin{displaymath}
  \det \left(\mathsf{L}-\mu\right)\; =\; -\mu^{3}\ +\
      \tfrac{1}{2}\mu\ \mathcal{C}_{(2)} \ +\
      \textstyle{1\over 3}\ \mathcal{C}_{(3)} \; ,
  \end{displaymath}
  and the fact that in the coloured case we must have $\mathcal{C}_{(k)}=0$.
}
\begin{equation}
\label{eq:96}
\mathsf{R}_{11}(\mu ) 
\; =\; 
-\frac{1}{\mu} \ -\ \frac{\psi_{1}}{2\mu^{2}}
\ -\ \frac{\psi_{1}\psi_{2}+f_{2}^{2}-\psi_{2}^{2}}{4\mu^{3}} \; .
\end{equation}
Comparing the two expressions we see that we must have
\begin{eqnarray}
  \label{eq:97}
  1 & =& 2xz+y^{2} \; , \\
  \label{eq:97b}
  \psi_{1} & =& 4xy \; , \\
  \label{eq:97c}
  4x^{2} & =& \psi_{1}\psi_{2}+f_{2}^{2}-\psi_{2}^{2} \; .
\end{eqnarray}
Eq.~(\ref{eq:97b}) was already obtained in eq.~(\ref{eq:79});
eq.~(\ref{eq:97}) leads to
\begin{equation}
  \label{eq:98}
  1 \; =\; \frac{2(r+a)^{2}\ +\ b}{2(r+a)^{2}\ +\ 2c}
  \hspace{.4cm}\mbox{whence}\hspace{.4cm}
  b\ =\ 2c \; .
\end{equation}
The l.h.s.~of eq.~(\ref{eq:97c}) appears in eq.~(\ref{eq:99}). Taking into
account that in the coloured case $\mathcal{C}_{(2)}=0$, the solution must be
such that
\begin{equation}
\label{eq:100}
f_{1}^{2}
\; = \; 
\psi_{1}^{2}-4x^{2}
\; = \; 
2\frac{(r+a)^{2}-c}{\left[  (r+a)^{2}+c \right]^{2}} \; .
\end{equation}
\par
Eliminating from the condition $\mathcal{C}_{(3)}=0$ the $f_{2}$ contribution
by means of eq.~(\ref{eq:97c}) we find
\begin{equation}
  \label{eq:101}
  \psi_{2} f_{1}^{2} \; =\; 4x^{2}\psi_{1}
  \hspace{.4cm}\xrightarrow{\;\;\;\mbox{whence}\;\;\;}\hspace{.4cm}
  \psi_{2}\; =\; -2\frac{r+a}{(r+a)^{2}-c} \; .
\end{equation}
At this point we know $\psi_{1}$ and $\psi_{2}$ so we can use
eq.~(\ref{eq:97c}) to find $f_{2}^{2}$:
\begin{equation}
  \label{eq:102}
  f_{2}^{2}\; =\; 2\frac{(r+a)^{2}+c}{\left[ (r+a)^{2}-c\right]^{2}} \; .
\end{equation}
If we now introduce the redefinitions $a=\lambda^{-1}$ and $c=\lambda^{-2}\xi$
then we find
\begin{figure}[t]
\centering
\includegraphics[height=8cm]{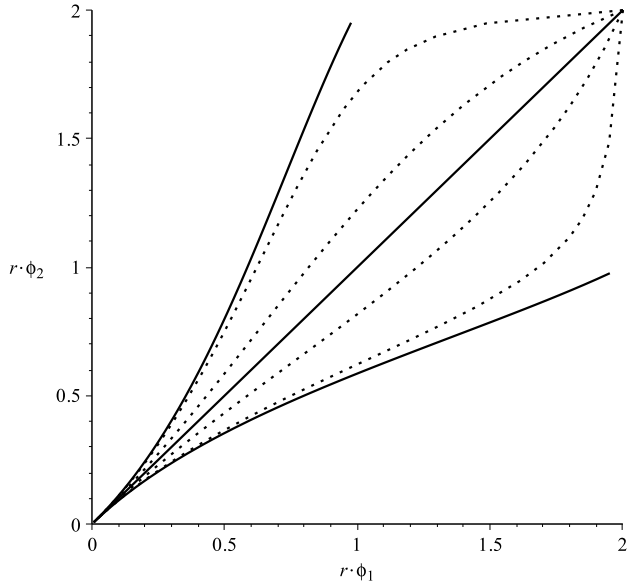}
\caption{\label{fig:FlowZeroCube} The flow described by the solutions in this
  section on the $(r\phi_{1},r\phi_{2})$-plane.  The solid line in the center
  corresponds to $\xi =0$, the upper solid line to $\xi =-1$ and the lower
  solid line to $\xi =1$.  The dashed lines correspond to various values of
  $|\xi |\neq 0,1$. All flows go towards the origin.  }
\end{figure}
\begin{equation}
  \label{eq:103}
  \begin{array}{lclclcl}
\psi_{1} 
& = & 
{\displaystyle 
-2\lambda\ \frac{1+\lambda r}{(1+\lambda r)^{2}\ +\ \xi}
}
&\;\;\; ,\;\;\; &
\psi_{2} 
& = & 
{\displaystyle 
-2\lambda\ \frac{1+\lambda r}{(1+\lambda r)^{2}\ -\ \xi}
}
 \; ,\\
& & & & & & \\
f_{1}^{2} 
& = & 
{\displaystyle
2\lambda^{2}\ \frac{(1+\lambda r)^{2}\ -\ \xi}{\left[ (1+\lambda r)^{2}\ +\
    \xi\right]^{2} }
}
& , &
f_{2}^{2} 
& = & 
{\displaystyle
2\lambda^{2}\ \frac{(1+\lambda r)^{2}\ +\ \xi}{\left[ (1+\lambda r)^{2}\ -\
    \xi\right]^{2} }
} \; .
  \end{array}
\end{equation}
For the solution to be well-defined on the range $r\in (0,\infty )$ we must
have $\lambda >0$ and $|\xi | \leq 1$.  
\par
The Higgs field is then readily seen to be
\begin{equation}
\label{eq:6}
\phi_{1}
\; =\; 
\frac{2\left( 1+\xi +\lambda r \right)}{r\left[ (1+\lambda r)^{2}+\xi\right]}
\hspace{.5cm},\hspace{.5cm}
\phi_{2}
\; =\; 
\frac{2\left( 1-\xi +\lambda r \right)}{r\left[ (1+\lambda r)^{2}-\xi\right]} \; .
\end{equation}
A special case is given by the solution with $\xi =0$: in that case the Higgs
field is given by
\begin{equation}
\label{eq:7}
\Phi 
\; =\; 
\frac{1}{r(1+\lambda r)}\ \mathrm{diag}(1, 0,-1) \; ,
\end{equation}
from which it is paramount that we are dealing with the embedding
$\mathrm{SU}(2)$'s coloured solution in eq.~(\ref{eq:4}), into
$\mathrm{SU}(3)$ by means of the maximal/singular embedding of
$\mathrm{SU}(2)$ into $\mathrm{SU}(3)$ \cite{Slansky:1981yr}.
\par
Figure~\ref{fig:FlowZeroCube} displays the behaviour of the above family of
solutions for the variables $r\phi_{i}$, which, if we take into account
eqs.~(\ref{eq:116}) and (\ref{eq:118}), together with the fact that the
solutions have no Higgs v.e.v., can be thought of as the evolution of the
colour charge along the flow parametrized by $r$.
\par
As promised at the beginning of this section, we now turn our attention to the
case in which there is a $2\times 2$ Jordan block in the spectral
representation of $\mathsf{L}$. We can choose vectors $\vec{v}$, $\vec{w}(2)$
and $\vec{w}(1)$ such that $\mathsf{L} \; =\; \vec{w}(2)\vec{w}(2)^{T}$ and
therefore, since the eigenvalues vanish in the defective case, 
\begin{equation}
\label{eq:8}
\mathsf{R}(\mu ) 
\; =\; 
-\frac{1}{\mu}
\left[
\vec{v}\vec{v}^{T}+\vec{w}(2)\vec{w}(1)^{T} +\vec{w}(1)\vec{w}(2)^{T}
\right]
 \ -\ 
\frac{1}{\mu^{2}}\ \vec{w}(2)\vec{w}(2)^{T} \; .
\end{equation}
The generic expression of the resolvent in eq.~(\ref{eq:96}), however, has a
$\mu^{-3}$ term whereas the chosen spectral representation does not: therefore
the coefficient of the $\mu^{-3}$ term must vanish, whence
$\psi_{1}\psi_{2}+f_{2}^{2}-\psi_{2}^{2}=0$. This, together with
eq.~(\ref{eq:99}), implies that $\psi_{1}^{2}=f_{1}^{2}$. Eq.~(\ref{eq:99b})
then implies that $\psi_{1}(\psi_{2}^{2}-f_{2}^{2})=0$. The conclusion then
must be that either $\psi_{1}=f_{1}=0$ and $\psi_{2}=\pm f_{2}$ or that
$\psi_{2}=f_{2}=0$ and $\psi_{1}=\pm f_{1}$; the non-zero functions are the
same as the $\mathrm{SU}(2)$ coloured solution in eq.~(\ref{eq:4}).
\par
That these solutions do not correspond to the regular embedding of
$\mathrm{SU}(2)$ into $\mathrm{SU}(3)$ becomes clear when we calculate the
Higgs field for {\em e.g.\/} the case $\psi_{1}=f_{1}=0$:
\begin{equation}
\label{eq:105}
2\Phi 
\; =\; 
\frac{1}{r}\ \mathrm{diag}(2,-1,-1) 
\; +\; 
\frac{1}{r(1+\lambda r)}\ \mathrm{diag}(0,1,-1) \; .
\end{equation}
This solution can be thought of as an $\mathrm{SU}(3)$ Wu-Yang monopole
coinciding with a coloured solution, a combination which is, however, not a
coloured solution even though it has no Higgs v.e.v. It does, however, show
the transmutation of colour charge between $r=0$ and $r=\infty$\footnote{In
  other words: the colour charge at spatial infinity and on the horizon are
  different and the solution smoothly interpolates between these two
  configurations.} typical of a coloured solution.
\section{$\mathcal{N}=2,d=4$ (Super-)EYM coloured black holes}
\label{sec:BHs}
The characterisation of supersymmetric solutions to $\mathcal{N}=2,d=4$
supergravity coupled to $m$ vector-multiplets with gauged isometries of
ref.~\cite{Hubscher:2008yz} called $\mathcal{N}=2,d=4$ (Super-)EYM does not
deal with the most general of such theories
\cite{Andrianopoli:1996cm,Freedman:2012zz}. In what follows it is important to
understand the restrictions; we will briefly outline the main ingredients and
restrictions, referring the reader to ref.~\cite{Hubscher:2008yz} for a more
detailed discussion.
\par
The bosonic field content of $\mathcal{N}=2,d=4$ supergravity coupled to $m$
vector multiplets consists of a metric, $m+1$ vector fields
$\mathsf{A}^{\Lambda}$ ($\Lambda =0,\ldots ,m$) and $m$ complex scalar fields
$\mathcal{Z}^{i}$ ($=1,\ldots ,m$). The scalar fields parametrise a
(special-)K\"ahler manifold $\mathcal{M}$ with a K\"ahler metric
$\mathcal{G}_{i\bar{\imath}}=\partial_{i}\partial_{\bar{\imath}}\mathcal{K}$,
where $\mathcal{K}$ is the K\"ahler potential. The basic premise of
$\mathcal{N}=2$ EYM is that the metric $\mathcal{G}$ allows for Killing
vectors and that we want to gauge a (necessarily non-Abelian!) subgroup
$\mathrm{G}\subseteq \mathrm{Isom}(\mathcal{G})$ whose associated Killing
vectors, denoted by $K_{\Lambda}$, satisfy
\begin{equation}
  \label{eq:25}
  \left[\ K_{\Lambda}\ ,\ K_{\Sigma}\ \right] 
\; =\; 
-f_{\Lambda\Sigma}{}^{\Omega}\ K_{\Omega} \; ,
\end{equation}
where the $f$'s are the structure constants of the Lie algebra $\mathfrak{g}$
associated to $\mathrm{G}$.  Observe that the way we are writing the Killing
vectors would mean that we are using $m+1$ isometries: this is the maximal
possibility and we can consider other possibilities by taking some
$K_{\Lambda}$ to vanish.
\par
The couplings of the scalars to themselves (that is, the K\"ahler metric) and
to the other fields are related and constrained by $\mathcal{N}=2$
supersymmetry and can be described in a unified way by a structure called
Special Geometry, see {\em e.g.\/}
\cite{deWit:1984px,Strominger:1990pd,Andrianopoli:1996cm}. The central object
in this description is a symplectic section $\mathcal{V}$ belonging to a flat
$2(m+1)$-dimensional bundle $E\times L^{1}\rightarrow \mathcal{M}$ with
structure group $\mathrm{Sp}(m+1;\mathbb{R})\times \mathrm{U}(1)$, that has to
satisfy
\begin{eqnarray}
  \label{eq:27}
  i & =& \langle\ \overline{\mathcal{V}}\ |\ \mathcal{V}\ \rangle \; ,\\
  \label{eq:27b}
 0 & =& \mathcal{D}_{\bar{\imath}}\mathcal{V} \; , \\ 
 \label{eq:27c}
 0 & =& \langle\ \mathcal{D}_{i}\mathcal{V}\ |\ \mathcal{V}\ \rangle \; ,
\end{eqnarray}
where the bracket denotes the $\mathrm{Sp}(m+1;\mathbb{R})$-invariant
innerproduct and the $\mathcal{D}$ denote $\mathrm{U}(1)$-covariant
(anti-)holomorphic derivatives. The action of the group we want to gauge can
then be lifted to the bundle and leads to the requirement that the symplectic
section $\mathcal{V}$ be invariant under $\mathrm{G}$ up to
$\mathrm{Sp}(m+1;\mathbb{R})$ and $\mathrm{U}(1)$ transformations. The
restriction imposed in ref.~\cite{Hubscher:2008yz} is then the statement that
the compensating $\mathrm{Sp}(m+1;\mathbb{R})$ transformations must be such
that under the branching to $\mathrm{G}$ only singlets and the adjoint
representation appears.
\par
The restriction is easier to understand in those situations where the so-called
preprotential $\mathcal{F}$ exists \cite{deWit:1984pk}: in that case we have a
homogeneous function $\mathcal{F}(\mathcal{X})$ of degree 2 such that
\begin{equation}
\label{eq:28}
\Omega 
= 
e^{-\mathcal{K}/2}\mathcal{V}=\left(\begin{array}{c} \mathcal{X}^{\Lambda}\\ 
\partial_{\Lambda}\mathcal{F}\end{array}\right)
  \hspace{.4cm}\longrightarrow\;\;\left\{
      \begin{array}{lcl}
             ie^{-\mathcal{K}} & =& \langle \overline{\Omega}|\Omega\rangle \; ,\\
             0 & =& \partial_{\bar{\imath}}\Omega \; , \\
             0 & =& \langle \partial_{i}\Omega |\Omega\rangle \; ,
      \end{array}
  \right.
\end{equation}
where $\mathcal{K}$ is the K\"ahler potential.  Ref.~\cite{Hubscher:2008yz}'s
restriction then means that the $\mathcal{X}$ transform under $\mathrm{G}$ as
the adjoint representation and singlets; this action must furthermore be such
that
\begin{equation}
\label{eq:26}
0 
\; =\; 
f_{\Lambda\Sigma}{}^{\Omega}\ \mathcal{X}^{\Sigma}\ \partial_{\Omega}\ \mathcal{F} \; ,
\end{equation}
which is the same thing as saying that the prepotential $\mathcal{F}$ is a
$\mathrm{G}$-invariant function \cite{deWit:1984pk}.
\par
The action of the bosonic sector of $\mathcal{N}=2$ super-EYM then reads (see {\em e.g.\/} \cite{Andrianopoli:1996cm,Freedman:2012zz})
\begin{equation}
\label{eq:N2d4SEYMaction}
  \begin{array}{rcl}
  S
  & = & 
{\displaystyle\int} d^{4}x \sqrt{|g|}
  \left[ R(g)
    +2\mathcal{G}_{i\bar{\jmath}}\mathsf{D}_{\mu}\mathcal{Z}^{i}\mathsf{D}^{\mu}\bar{\mathcal{Z}}^{\bar{\jmath}}
    +2\mathrm{Im}\left(\mathcal{N}\right)_{\Lambda\Sigma} \
    \mathsf{F}^{\Lambda\, \mu\nu}\ \mathsf{F}^{\Sigma}{}_{\mu\nu}
  \right. \\
  & & \\
  & & \left. 
    -2\mathrm{Re}\left(\mathcal{N}\right)_{\Lambda\Sigma}  
    \mathsf{F}^{\Lambda\, \mu\nu}\ \star \mathsf{F}^{\Sigma}{}_{\mu\nu}
    \ -\ V(Z,Z^{*})
  \right]\, ,
\end{array}
\end{equation}
where the gauge-covariant derivative and field strengths are defined as
\begin{equation}
\mathsf{D} \mathcal{Z}^{i} \ =\ d\mathcal{Z}^{i}\ +\ \mathsf{A}^{\Lambda}\ K_{\Lambda}{}^{i}
\hspace{.5cm}\mbox{and}\hspace{.5cm}
\mathsf{F}^{\Lambda}\ =\ d\mathsf{A}^{\Lambda} \ -\ \textstyle{1\over 2}\ f_{\Sigma\Gamma}{}^{\Lambda}\ \mathsf{A}^{\Sigma}\wedge\mathsf{A}^{\Gamma} \; .
\end{equation}
$\mathcal{N}_{\Lambda\Sigma}$ is the, model dependent, period matrix and $V(Z,Z^{*})$ is
the scalar potential\footnote{
 Observe that since the imaginary part of the period matrix is negative definite,
 the scalar potential is positive semidefinite.
}
\begin{equation}
V(\mathcal{Z},\bar{\mathcal{Z}}) \ =\
-{\textstyle\frac{1}{4}}\
\mathrm{Im}\left(\mathcal{N}\right)^{\Lambda\Sigma}\ \mathcal{P}_{\Lambda}\mathcal{P}_{\Sigma}\, .
\end{equation}
In the above equations, $K_{\Lambda}{}^{i}(\mathcal{Z})$ are the holomorphic
Killing vectors of the isometries that have been gauged
and $\mathcal{P}_{\Lambda}(\mathcal{Z},\bar{\mathcal{Z}})$ are the corresponding momentum maps;
they are related by
\begin{equation}
  \label{eq:33}
  K_{\Lambda}^{i} \; =\; i\ \mathcal{G}^{i\bar{\jmath}}\ \partial_{\bar{\jmath}}\mathcal{P}_{\Lambda} \; .
\end{equation}
\par
Once we have defined the model, one sees that the supersymmetric solutions in
the so-called timelike case which is the one the black holes belong to, is
constructed out of $2(m+1)$ real seed functions denoted by
$\mathcal{I}^{\Lambda}$ and $\mathcal{I}_{\Lambda}$.  These seed functions are
such that if the $\Lambda$ index corresponds to a singlet under the
$\mathrm{G}$-action, the corresponding $\mathcal{I}^{\Lambda}$ and
$\mathcal{I}_{\Lambda}$ are harmonic functions on $\mathbb{R}^{3}$, whereas if
it corresponds to the adjoint, the Higgs field defined by  
\begin{equation}
\mathcal{I}^{\Lambda}
\; = \; 
-\sqrt{2}\Phi^{\Lambda}\; ,
\end{equation}
must solve the Bogomol'nyi equation (\ref{eq:1}). The corresponding
$\mathcal{I}_{\Lambda}$ must solve an involved equation, see
\cite[eq.~(4.27)]{Hubscher:2008yz}, and we will avoid this hurdle by taking
them to vanish identically, which is always possible. Observe, however, that
this means that the non-Abelian gauge fields will have no electric components
and will be purely magnetic.
\par
From this point onwards, the construction of the solution is basically the
same as for an ungauged Abelian theory \cite{Behrndt:1997ny,Meessen:2006tu}:
as we are interested in spherically symmetric, static spacetimes we must
impose the condition that $\langle\mathcal{I}|\mathsf{D}\mathcal{I}\rangle
=0$, where $\mathsf{D}$ is the $\mathrm{G}$-covariant derivative.  Having
satisfied said condition, the metric becomes
\begin{equation}
  \label{eq:MetDef}
  ds^{2}\; =\; e^{2U}dt^{2}\; -\; e^{-2U}\ d\vec{x}_{(3)}^{2} 
  \hspace{1cm}\mbox{where}\hspace{1cm}
  e^{-2U} \; =\; \mathrm{W}(\mathcal{I}) \; ,
\end{equation}
is a model-dependent homogeneous function of degree 2 called the Hesse
potential \cite{Bates:2003vx,Mohaupt:2011aa,Meessen:2011aa}.  The scalar
fields are model-dependent but they are nicely expressed in terms of the Hesse
potential as
\begin{equation}
  \label{eq:29}
Z^{i} 
\; =\; 
\frac{\tilde{\mathcal{I}}^{i}\ +\ i \mathcal{I}^{i}}{\tilde{\mathcal{I}}^{0}
\ +\ i\mathcal{I}^{0}}
\hspace{2cm}\mbox{where}\hspace{1cm}
\tilde{\mathcal{I}}^{\Lambda}
\; =\; 
-\tfrac{1}{2}\ \frac{\partial\mathrm{W}}{\partial \mathcal{I}_{\Lambda}} \; .
\end{equation}
\par
The final step in the construction consists in adjusting the integration
constants of the seed functions so that the solution describes the geometry of
an extremal black hole outside its horizon: in order for a spherically
symmetric metric of the type in eq.~(\ref{eq:MetDef}) to be interpreted as a
black hole, it should be asymptotically flat and singular at $r=0$ only and
this singularity must be such that the limiting geometry at $r=0$ is that of
an $aDS_{2}\times S^{2}$ spacetime. With these provisos, the area of the
2-sphere in this limiting geometry corresponds to the area of the horizon of
the black hole and therefore also to the entropy of the black hole (See {\em
  e.g.\/} \cite{Ortin:2004ms}).  For a given Hesse potential and given seed
functions, this entropy can be calculated straightforwardly but need not be
finite or positive: the regularity of the metric at the horizon is then, by
means of the entropy, a constraint on the parameters of the solution.
\par
Further constraints on the parameters of the seed functions come from the fact
that the metrical factor $e^{-2U}$ must never vanish on the interval $r\in
(0,\infty )$, as otherwise the solution would have a  curvature singularity on
said interval, whence ruining the interpretation of the solution as describing
the exterior of an extremal black hole.
\par
Having outlined the restrictions on the theories and the solutions, we are
ready to use the coloured solutions to the $\mathrm{SU}(3)$ Bogomol'nyi
equations obtained in sec.~\ref{sec:DefSU3} to build supersymmetric coloured
black holes.
\subsection{$\overline{\mathbb{CP}}^{8}$ coloured black holes}
\label{sec:CP8}
The so-called $\overline{\mathbb{CP}}^{8}$ model has 9 vector fields, a scalar
manifold that is the symmetric space $\mathrm{SU}(1,8)/\mathrm{U}(8)$ and is
defined by the prepotential
\begin{equation}
\label{eq:18}
\mathcal{F}
\; =\; 
\textstyle{1\over 4i}\ \eta_{\Lambda\Sigma}\ \mathcal{X}^{\Lambda}\mathcal{X}^{\Sigma}
\hspace{2cm}\mbox{where}\;\; 
\eta 
\ =\ 
\mathrm{diag}(+,[-]^{8}) \; ,
\end{equation}
and the indices $\Lambda$ and $\Sigma$ run from $0$ to $8$. Since the
prepotential is manifestly $\mathrm{SO}(1,8)$ invariant and we have 9 vector
fields at our disposal we can at most gauge a 9-dimensional subgroup
$\mathrm{G}\subset \mathrm{SO}(1,8)$; if we couple this to the restriction
that in the branching of $\mathrm{SO}(1,8)$'s $\mathbf{9}$ we must only find
the adjoint and singlets of $\mathrm{G}$, then we see that the singular
embedding of $\mathrm{SU}(3)$ into $\mathrm{SO}(8)$ does the trick
\cite{Slansky:1981yr}.
\par
Sticking to a purely magnetic solution, so that $\mathcal{I}_{0}=0$, we see
that the metrical factor becomes
\begin{equation}
  \label{eq:19}
  e^{-2U} \; =\; \tfrac{1}{2} \left(I^{0}\right)^{2} \ -\ \tfrac{1}{2}I^{a}I^{a}
           \; =\; \tfrac{1}{2} \left(I^{0}\right)^{2} \ -\ 2\mathrm{Tr}\left(\Phi^{2}\right)
           \; =\; H^{2}\ -\ \phi_{1}^{2}+\phi_{1}\phi_{2}\ -\ \phi_{2}^{2} \; ,
\end{equation}
where in the last step we redefined $I^{0}=\sqrt{2}H$. Since $H$ is a
$\mathrm{SU}(3)$-singlet, it has to be a harmonic function and, to preserve
spherical symmetry, it has to be of the form
\begin{equation}
  \label{eq:20}
  H\; =\; h\ +\ \frac{p}{r}
  \hspace{1cm}\mbox{so that}\hspace{1cm}
  \lim_{r\rightarrow\infty} e^{2U}\; =\; h^{2}\, +\, \frac{2hp}{r}\, +\ldots
\end{equation}
where we have already used the asymptotic vanishing of the Higgs field
characteristic of coloured solutions. We take $h^{2}=1$ in order for the
metric to become asymptotically Minkowski in spherical coordinates.  The mass
is then given by $M=hp$, which can always be taken to be positive by taking
$h=\mathrm{sign}(p)$, whence $M=|p|$.  The entropy of the non-Abelian black
hole is readily calculated to be
\begin{equation}
  \label{eq:21}
   \frac{S}{\pi} \; =\; \left\{
           \begin{array}{lclcl}
             p^{2}\ -\ 4  &\hspace{.4cm}:\hspace{.4cm} &  \xi & \neq& \pm 1 \\
                & & & & \\
             p^{2}\ -\ 3  & : & \xi & =& \pm 1
           \end{array}
      \right.
\end{equation}
where the $4$ and the $3$ are the monopole's contribution to the entropy.  We
must have $|p|>2$ when $\xi\neq 1$ and $|p|>\sqrt{3}$ when $\xi = \pm 1$ in
order for the entropy to be finite and positive.  As
in the ungauged case, positive mass and a well-defined entropy of the horizon
is, at least for these solutions, enough to ensure the regularity of the
metric.
\par
The eight complex physical scalars in this theory, $\mathcal{Z}^{a}$
($a=1,\ldots ,8$) are given by
\begin{equation}
  \label{eq:22}
  \mathcal{Z}^{a}\; =\; \frac{I^{a}}{I^{0}} \; =\; -\frac{\Phi^{a}}{H} 
  \hspace{.4cm}
\xrightarrow{\;\;\mbox{$\mathrm{SU}(3)$ defining rep.}\;\;}
\hspace{.4cm}
  \mathcal{Z}\; =\; -\frac{\Phi}{H} \; .
\end{equation}
In the last expression it is paramount that the scalars behave asymptotically
as $\mathcal{Z}\sim \mathcal{O}(r^{-2})$ and that near the horizon they behave
as
\begin{equation}
  \label{eq:24}
  \lim_{r\rightarrow 0}\mathcal{Z} \; =\; \left\{
         \begin{array}{lclcl}
                -{\displaystyle\frac{1}{p}}\ \mathrm{diag}(1,0,-1)   &\;\;\; :\;\;\; & \xi & \neq & \pm 1 \\
                    & & & &  \\
                -{\displaystyle\frac{1}{2p}}\ \mathrm{diag}(2,-1,-1) & : & \xi & =& +1 \\
                    & & & &  \\
                -{\displaystyle\frac{1}{2p}}\ \mathrm{diag}(1,1,-2) & : & \xi & =& -1
         \end{array}
         \right.
\end{equation}
Let us stress once again that even though the near-horizon behaviour is
determined by (colour) charges only the Abelian charge $p$ is asymptotically
measurable. Also observe that the hair parameters $\lambda$ and $\xi$ do not
influence the asymptotic nor the near-horizon behaviour, illustrating once
again the non-applicability of the no-hair theorem in gravity coupled to YM
theories.
\subsection{$\mathbb{C}$-magic coloured black holes}
\label{sec:CMagic}
The $\mathbb{C}$-magic model is a model with 10 vector fields and 9 complex
scalars parametrising the symmetric space $\mathrm{SU}(3,3)/S\left[
  \mathrm{U}(3)\times \mathrm{U}(3)\right]$.  A convenient prepotential for
this model was given in ref.~\cite{Ferrara:2006yb} and splits the 10 complex
coordinates $\mathcal{X}^{\Lambda}$ into a singlet under the
$\mathrm{SU}(3)\times\mathrm{SU}(3)$ action, $\mathcal{X}^{0}$, and a $3\times
3$ matrix that we denote by $\mathcal{X}$, transforming as the
$(\mathbf{3},\bar{\mathbf{3}})$ irrep. The prepotential can then be expressed
as
\begin{equation}
  \label{eq:23}
  \mathcal{F}
\; = \; 
\frac{\det\left(\mathcal{X}\right)}{\mathcal{X}^{0}} \; .
\end{equation}
As was argued in ref.~\cite{Hubscher:2008yz} this model allows for a gauging
of the diagonal $\mathrm{SU}(3)$, which seeing the branching rule
$(\mathbf{3},\bar{\mathbf{3}})\rightarrow \mathbf{1}\oplus\mathbf{8}$
\cite{Slansky:1981yr} is exactly what is needed to fulfill the constraints
outlined in sec.~\ref{sec:BHs}.
\par
Ref.~\cite{Hubscher:2008yz} then used the spherically symmetric Wilkinson-Bais
monopoles \cite{Wilkinson:1978zh} for $\mathrm{SU}(3)$ to construct
supersymmetric globally regular solutions describing the (super-)gravitational
backreaction of said monopoles and also used a small generalisation to
construct supersymmetric non-Abelian black hole. All those solutions have,
however, asymptotic colour charge and non-vanishing Higgs v.e.v.~and in this
section we are going to use the solutions of the Bogomol'nyi equation derived
in sec.~\ref{sec:DefSU3} to construct coloured black-hole solutions to the
$\mathbb{C}$-magic model.
\par
Following ref~\cite{Hubscher:2008yz}, we shall take $\mathcal{I}^{0}=0$, which
means that the Abelian gauge field $\mathsf{A}^{0}$ is purely electric, and
also $\mathcal{I}_{\Lambda\neq 0}=0$. This last choice leads to a static
solution with a metric factor with the simple form
\begin{equation}
\label{eq:9}
 e^{-2U} 
\; =\; 
\sqrt{\ H\ \det\left( K\ \mathrm{Id}-2\Phi\right)  \ }
\; =\; 
\sqrt{\ H\ (K-\phi_{1})(K-\phi_{2}+\phi_{1})(K+\phi_{2})\ } \; ,
\end{equation}
where we defined $\mathcal{I}_{0}=\textstyle{1\over 4\sqrt{2}}\ H$ and took
the singlet seed function to be $\textstyle{1\over
  \sqrt{2}}K$.\footnote{Observe that ref.~\cite{Hubscher:2008yz} uses the
  definition $\mathcal{I}_{0}=\textstyle{1\over \sqrt{2}}\ H$, which means
  that the metrical factor in \cite[eq.~(5.47)]{Hubscher:2008yz} is missing a
  factor of two.  In said equation, the harmonic function $K$ is denoted by
  $\lambda$, a naming we changed here in order to avoid confusion.}  As $H$
and $K$ are singlets under the gauge group, they are harmonic functions and
taking them to be spherically symmetric they can be expanded as
\begin{equation}
  \label{eq:10}
  H\; =\; h\ +\ \frac{q}{r} \;\;\;\; ,\;\;\;
  K\; =\; k\ +\ \frac{p}{r} \; .
\end{equation}
The criterion for the absence of coordinate singularities for any $r\in
(0,\infty )$ immediately implies that $\mathrm{sign}(h)=\mathrm{sign}(q)$.
\par
By considering the asymptotic behaviour of the metrical factor in
eq.~(\ref{eq:9}) we can straightforwardly normalise the solution to asymptote
to ordinary Minkowski spacetime by taking $hk^{3}=1$ which then leads to the
following expression for the mass
\begin{equation}
  \label{eq:11}
  M\; =\; \frac{k^{3}q\ +\ 3k^{-1}p}{4} \; .
\end{equation}
$k$ is related to the values of the scalars at spatial infinity. The actual
expression is complicated and not very enlightening.
\par
As one can see from Figure~\ref{fig:FlowZeroCube} the limit $r\rightarrow 0$
differs substantially between the cases $\xi \neq \pm 1$ and $\xi =\pm 1$ and
we will discuss the regularity properties of the solution seperately.  From
the metrical factor in eq.~(\ref{eq:11}) it follows that the case $\xi =-1$
can be obtained from the case $\xi =1$ by substituting
$\phi_{1}\leftrightarrow \phi_{2}$, $K\rightarrow -K$ and $H\rightarrow -H$,
and, accordingly, we shall not discuss the construction of a coloured
black-hole solution for the case $\xi =-1$.
\subsubsection{$\mathbf{\xi\neq \pm 1}$}
In this case the entropy reads
\begin{equation}
  \label{eq:12}
  S(\xi \neq \pm 1) \; =\; \pi\ \sqrt{\ qp(p^{2}-4)\ } \; .
\end{equation}
there are four possible conditions on the charges that make the entropy well
defined and finite:
\begin{itemize}
\item[Case a)] $p>2$ and $q>0$. The absence of zeroes in the function $H$ then
  implies that $h>0$, whence also $k>0$ by the normalisation condition; The
  mass as calculated by eq.~(\ref{eq:11}) is automatically positive.
\item[Case b)] $p<-2$ and $q<0$. This further implies $h<0$ and $k<0$ implying
  that the mass is automatically positive.
\item[Case c)] $p\in (0,2)$ and $q<0$. Seeing that we must have $h<0$ and
  $k<0$, the mass is not automatically positive.  One can, however, see fairly
  rapidly that in this case the combination $H-\phi_{2}$ appearing in the
  metrical factor (\ref{eq:9}) has a zero on the interval $(0,\infty )$:
  conforming to the criteria outlined above, this means that case c) is not a
  viable option and must be discarded.
\item[Case d)] $p\in (-2,0)$ and $q>0$. This case must be supplemented by the
  conditions $h>0$ and $k>0$, which then means that the factor $H-\phi_{1}$
  has a zero, whence this case must also be discarded.
\end{itemize}
For the cases a) and b) one can see that the resulting metrical factor has no
zeroes for $r>0$ and the corresponding solutions describe coloured black
holes.
\subsubsection{$\mathbf{\xi =  1}$}
The entropy reads
\begin{equation}
  \label{eq:13}
  S(\xi = 1) \; =\;  \pi\ \sqrt{\ q(p + 1)^{2}(p - 2)\ } \; .
\end{equation}
There are possibilities:
\begin{itemize}
\item[Case $\alpha$)] $p>2$ and $q>0$, which following the same reasoning as
  before leads to the further constraints $h>0$ and $k>0$ and therefore also
  to positive mass.
\item[Case $\beta$)] $p<-1$ and $q<0$, whence also $h<0$ and $k<0$, which also
  implies that the mass is positive.
\end{itemize}
In these two cases the metrical factor is free of coordinate singularities on
the interval $(0,\infty )$ and do define coloured black holes.
\par
Observe that as far as the entropy is concerned, we could allow for the
possibility of $p\in (-1,2)$. This possibility requires $q<0$, $h<0$ and
$k<0$, which automatically implies that the factor $H+\phi_{2}$ in the
metrical factor has a zero and must therefore be discarded.
\section{Conclusions}
\label{sec:Concl}
In this article we have constructed coloured black hole solutions to two
models of $\mathcal{N}=2,d=4$ EYM theories with an $\mathrm{SU}(3)$ gauge
group, namely the $\overline{\mathbb{CP}}^{8}$ and the $\mathbb{C}$-magic
model.  This construction was possible due to the fact that the Bogomol'nyi
equation, a prominent ingredient in the construction of supersymmetric
solutions to the used class of theories, under the assumption of (maximal)
spherical symmetry is an integrable system. The system admits a Lax pair and
after having identified the solutions needed to construct coloured black holes
as corresponding to defective Lax matrices, these were constructed for the
$\mathrm{SU}(3)$ Bogomol'nyi equation. The coloured black holes built upon
these solutions have the same characteristics as the $\mathrm{SU}(2)$ black
holes: there is no asymptotic colour charge, one always needs extra active
Abelian fields and most importantly of all the horizon is colourful and the
entropy depends only on the colour charge on the horizon and not on the hair
parameters of the solutions.
\par
The transmutation solution in eq.~(\ref{eq:105}) leads to a type of black hole
that is unavailable in the $\mathrm{SU}(2)$-gauged models considered up till
now: as one can see, the solution is such that the asymptotic colour charge is
non-vanishing but different from the horizon colour charge and it can be used
to construct black hole solutions along the lines outlined in the last
section.  Consider for example the case of $\overline{\mathbb{CP}}^{8}$
treated in sec.~\ref{sec:CP8}, the embedding into the $\mathbb{C}$-magic
model being also straightforward: the metrical factor can be expanded as
\begin{equation}
\label{eq:31}
e^{-2U}
\; =\; 
h^{2}
\ +\ \frac{2hp}{r}
\ +\ \frac{p^{2}-4}{r^{2}}
\ +\ \frac{2\lambda}{r(1+\lambda r)^{2}} \; .
\end{equation}
According to the criteria outlined in sec.~\ref{sec:BHs} we see that in order
to have a regular black-hole solution we must have $h=\mathrm{sign}(p)$ so
that the mass is $M=|p|\geq 0$ and $|p|>2$ so that the entropy, given by $S=
\pi (p^{2}-4)$ is finite and positive. These two conditions suffice to make
the metrical factor a sum of positive terms, producing a perfectly
well-defined black-hole solution. By looking at the expression for the scalars
in eq.~(\ref{eq:24}) one can readily see that they are also well defined and
behave near the horizon as the $\xi\neq \pm 1$ case in eq.~(\ref{eq:24}).
\par
In sec.~\ref{sec:DefSU3} we saw that the $\mathrm{SU}(3)$ coloured solution
is related to a $3\times 3$ Jordan block, whereas the transmutation solution
is related to the appearance of a smaller Jordan block. In fact, it can be
shown that the $\mathrm{SU}(N)$ coloured solutions can only follow from a Lax
matrix that is similar to an $N\times N$ Jordan block, meaning that the
transmutation solutions are the rule rather the exception for $N>3$.  The
question of whether these transmutation solutions can, like the one for
$\mathrm{SU}(3)$, be written as the sum of Wu-Yang monopoles and coloured
solutions for smaller $N$ should be worth investigating.
\par
The coloured and the transmutation black holes have the characteristic that
the asymptotic colour charge does not match the colour charge seen on the
horizon. Strictly speaking, non-Abelian charges can only be defined globally,
at infinity but the calculation of the entropy indicates clearly the presence
of non-Abelian charges, different from those seen at spatial infinity, which
contribute to it. This is a very intriguing phenomenon which calls for further
investigation. A microscopic interpretation of the entropy of non-Abelian
black holes (coloured or with asymptotic charges) is badly needed. 
\par 
As for the celebrated attractor mechanism
\cite{Ferrara:1995ih,Strominger:1996kf,Ferrara:1996dd,Ferrara:1996um}, it
works (in the covariant sense discovered in ref.~\cite{Huebscher:2007hj}), but
only in terms of the horizon charges. A further difference from the well-known
Abelian case is the fact that the asymptotic value of the scalars related to
the Higgs field is not arbitrary: it has to vanish for coloured black
holes, as discussed in sec.~\ref{sec:SpherBogo}. 
\par
We hope to have convinced the reader that the physics of non-Abelian charged
black holes has very interesting features that go far beyond the well-known
existence of non-Abelian hair and deserves further investigation.  Work in
this direction is in progress.
\section*{Acknowledgments}
PM wishes to thank the Instituto de F\'{\i}sica Te\'orica UAM/CSIC for its
continued hospitality.  This work has been supported in part by the Spanish
Ministry of Science and Education grant FPA2012-35043-C02 (-01 {\&} -02), the
Centro de Excelencia Severo Ochoa Program grant SEV-2012-0249, the Comunidad
de Madrid grant HEPHACOS S2009ESP-1473, EU-COST action MP1210 ``The String
Theory Universe' and the Ram\'on y Cajal fellowship RYC-2009-0501.  TO wishes
to thank M.M.~Fern\'andez for her unfaltering support.
\bibliographystyle{pet} 
\bibliography{pair}

\begin{thebibliography}{10}
\newcommand{\enquote}[1]{``#1''}
\expandafter\ifx\csname url\endcsname\relax
  \def\url#1{\texttt{#1}}\fi
\expandafter\ifx\csname urlprefix\endcsname\relax\def\urlprefix{URL }\fi
\providecommand{\eprint}[2][]{\url{#2}}

\bibitem{Huebscher:2007hj}
M.~H{\"u}bscher, P.~Meessen, T.~Ort\'{\i}n, S.~Vaul\`a:
  \emph{\enquote{{Supersymmetric N=2 Einstein-Yang-Mills monopoles and
  covariant attractors}}, }Phys.~Rev. \textbf{D78}(2008), 065031.
  (\texttt{0712.1530}).

\bibitem{Hubscher:2008yz}
--- \emph{\enquote{{N=2 Einstein-Yang-Mills's BPS solutions}}, }JHEP
  \textbf{0809}(2008), 099. (\texttt{0806.1477}).

\bibitem{deWit:1984px}
B.~de~Wit, P.~Lauwers, A.~Van~Proeyen: \emph{\enquote{{Lagrangians of N=2
  supergravity-matter Systems}}, }Nucl.~Phys. \textbf{B255}(1985), 569.

\bibitem{Andrianopoli:1996cm}
L.~Andrianopoli, M.~Bertolini, A.~Ceresole, R.~D'Auria, S.~Ferrara, P.~Fre,
  T.~Magri: \emph{\enquote{{N=2 supergravity and N=2 superYang-Mills theory on
  general scalar manifolds: Symplectic covariance, gaugings and the momentum
  map}}, }J.~Geom.~Phys. \textbf{23}(1997), 111--189.
  (\texttt{hep-th/9605032}).

\bibitem{Behrndt:1997ny}
K.~Behrndt, D.~L{\"u}st, W.~A. Sabra: \emph{\enquote{{Stationary solutions of
  N=2 supergravity}}, }Nucl.~Phys. \textbf{B510}(1998), 264--288.
  (\texttt{hep-th/9705169}).

\bibitem{Meessen:2006tu}
P.~Meessen, T.~Ort\'{\i}n: \emph{\enquote{{The Supersymmetric configurations of
  N=2, D=4 supergravity coupled to vector supermultiplets}}, }Nucl.~Phys.
  \textbf{B749}(2006), 291--324. (\texttt{hep-th/0603099}).

\bibitem{Meessen:2008kb}
P.~Meessen: \emph{\enquote{{Supersymmetric coloured/hairy black holes}},
  }Phys.~Lett. \textbf{B665}(2008), 388--391. (\texttt{0803.0684}).

\bibitem{Bueno:2014mea}
P.~Bueno, P.~Meessen, T.~Ort\'{\i}n, P.~F. Ram\'{\i}rez: \emph{\enquote{{N=2
  Einstein-Yang-Mills's static two-center solutions}}, }{JHEP}
  \textbf{1412}(2014), 093. (\texttt{1410.4160}).

\bibitem{Bartnik:1988am}
R.~Bartnik, J.~McKinnon: \emph{\enquote{{Particle-like solutions of the
  Einstein Yang-Mills equations}}, }Phys.~Rev.~Lett. \textbf{61}(1988),
  141--144.

\bibitem{Bizon:1990sr}
P.~Bizon: \emph{\enquote{{Colored black holes}}, }Phys.~Rev.~Lett.
  \textbf{64}(1990), 2844--2847.

\bibitem{Kuenzle:1990is}
H.~K{\"u}nzle, A.~Masood-ul Alam: \emph{\enquote{{Spherically symmetric static
  $\mathrm{SU}(2)$ Einstein Yang-Mills fields}}, }J.~Math.~Phys.
  \textbf{31}(1990), 928--935.

\bibitem{Volkov:1990sva}
M.~Volkov, D.~Gal'tsov: \emph{\enquote{{Black holes in Einstein Yang-Mills
  theory. (In Russian)}}, }Sov.~J.~Nucl.~Phys. \textbf{51}(1990), 747--753.

\bibitem{Fan:2014ixa}
Z.-Y. Fan, H.~Lü: \emph{\enquote{{SU(2)-Colored (A)dS Black Holes in Conformal
  Gravity}}, }JHEP \textbf{1502}(2015), 013. (\texttt{1411.5372}).

\bibitem{Leznov:1980tz}
A.~Leznov, M.~Saveliev: \emph{\enquote{{Representation Theory and Integration
  of Nonlinear Spherically Symmetric Equations to Gauge Theories}},
  }Commun.~Math.~Phys. \textbf{74}(1980), 111--118.

\bibitem{Koikawa:1981xg}
T.~Koikawa: \emph{\enquote{{Exact $\mathrm{SU}(N)$ monopole solutions with
  spherical symmetry by the inverse scattering method}}, }Phys.~Lett.
  \textbf{B110}(1982), 129.

\bibitem{Farwell:1982du}
R.~Farwell, M.~Minami: \emph{\enquote{{One-dimensional Toda molecule 1: general
  solution}}, }Prog.~Theor.~Phys. \textbf{69}(1983), 1091.

\bibitem{Farwell:1983sx}
--- \emph{\enquote{{One-dimensional Toda molecule 2: the solutions applied to
  Bogomolny monopoles with spherical symmetry}}, }Prog.~Theor.~Phys.
  \textbf{70}(1983), 710.

\bibitem{Anderson:1995sz}
A.~Anderson: \emph{\enquote{{An elegant solution of the N body Toda problem}},
  }J.~Math.~Phys. \textbf{37}(1996), 1349--1355. (\texttt{hep-th/9507092}).

\bibitem{Lu:1996jr}
H.~Lu, C.~Pope: \emph{\enquote{{SL(N+1,R) Toda solitons in supergravities}},
  }Int.~J.~Mod.~Phys. \textbf{A12}(1997), 2061--2074.
  (\texttt{hep-th/9607027}).

\bibitem{Fre:2009dg}
P.~Fre, A.~S. Sorin: \emph{\enquote{{The integration algorithm for nilpotent
  orbits of G/H* Lax systems: for extremal black holes}}, } (2009),.
  (\texttt{0903.3771}).

\bibitem{Chemissany:2009af}
W.~Chemissany, P.~Fre, A.~S. Sorin: \emph{\enquote{{The integration algorithm
  of Lax equation for both generic Lax matrices and generic initial
  conditions}}, }Nucl.~Phys. \textbf{B833}(2010), 220--225.
  (\texttt{0904.0801}).

\bibitem{Chemissany:2010zp}
W.~Chemissany, P.~Fre, J.~Rosseel, A.~Sorin, M.~Trigiante, \emph{et~al.}:
  \emph{\enquote{{Black holes in supergravity and integrability}}, }JHEP
  \textbf{1009}(2010), 080. (\texttt{1007.3209}).

\bibitem{zbMATH03141695}
H.-C. {Wang}: \emph{\enquote{{On invariant connections over a principal fibre
  bundle.}}, }{Nagoya Math.~J.} \textbf{13}(1958), 1--19.

\bibitem{Wilkinson:1978zh}
D.~Wilkinson, F.~A. Bais: \emph{\enquote{{Exact $\mathrm{SU}(N)$ monopole
  solutions with spherical symmetry}}, }Phys.~Rev. \textbf{D19}(1979), 2410.

\bibitem{Ganoulis:1981sx}
N.~Ganoulis, P.~Goddard, D.~I. Olive: \emph{\enquote{{Selfdual Monopoles and
  Toda Molecules}}, }Nucl.~Phys. \textbf{B205}(1982), 601.

\bibitem{Kuenzle:1991wa}
H.~K{\"u}nzle: \emph{\enquote{{$\mathrm{SU}(n)$ Einstein Yang-Mills fields with
  spherical symmetry}}, }Class.~Quant.~Grav. \textbf{8}(1991), 2283--2297.

\bibitem{Protogenov:1977tq}
A.~Protogenov: \emph{\enquote{{Exact classical solutions of Yang-Mills
  sourceless equations}}, }Phys.~Lett. \textbf{B67}(1977), 62--64.

\bibitem{Slansky:1981yr}
R.~Slansky: \emph{\enquote{{Group theory for unified model building}},
  }Phys.~Rept. \textbf{79}(1981), 1--128.

\bibitem{Freedman:2012zz}
D.~Z. Freedman, A.~Van~Proeyen: \emph{{Supergravity}}; Cambridge U.P., 2012.

\bibitem{Strominger:1990pd}
A.~Strominger: \emph{\enquote{{Special geometry}}, }Commun.~Math.~Phys.
  \textbf{133}(1990), 163--180.

\bibitem{deWit:1984pk}
B.~de~Wit, A.~Van~Proeyen: \emph{\enquote{{Potentials and symmetries of general
  gauged N=2 supergravity: Yang-Mills models}}, }Nucl.~Phys.
  \textbf{B245}(1984), 89.

\bibitem{Bates:2003vx}
B.~Bates, F.~Denef: \emph{\enquote{{Exact solutions for supersymmetric
  stationary black hole composites}}, }JHEP \textbf{1111}(2011), 127.
  (\texttt{hep-th/0304094}).

\bibitem{Mohaupt:2011aa}
T.~Mohaupt, O.~Vaughan: \emph{\enquote{{The Hesse potential, the c-map and
  black hole solutions}}, }JHEP \textbf{1207}(2012), 163. (\texttt{1112.2876}).

\bibitem{Meessen:2011aa}
P.~Meessen, T.~Ort\'{\i}n, J.~Perz, C.~Shahbazi: \emph{\enquote{{H-FGK
  formalism for black-hole solutions of N=2, d=4 and d=5 supergravity}},
  }Phys.~Lett. \textbf{B709}(2012), 260--265. (\texttt{1112.3332}).

\bibitem{Ortin:2004ms}
T.~Ort\'{\i}n: \emph{{Gravity and strings}}; Cambridge U.P., 2004.

\bibitem{Ferrara:2006yb}
S.~Ferrara, E.~G. Gimon, R.~Kallosh: \emph{\enquote{{Magic supergravities, N= 8
  and black hole composites}}, }Phys.~Rev. \textbf{D74}(2006), 125018.
  (\texttt{hep-th/0606211}).

\bibitem{Ferrara:1995ih}
S.~Ferrara, R.~Kallosh, A.~Strominger: \emph{\enquote{{N=2 extremal black
  holes}}, }Phys.~Rev. \textbf{D52}(1995), 5412--5416.
  (\texttt{hep-th/9508072}).

\bibitem{Strominger:1996kf}
A.~Strominger: \emph{\enquote{{Macroscopic entropy of N=2 extremal black
  holes}}, }Phys.~Lett. \textbf{B383}(1996), 39--43. (\texttt{hep-th/9602111}).

\bibitem{Ferrara:1996dd}
S.~Ferrara, R.~Kallosh: \emph{\enquote{{Supersymmetry and attractors}},
  }Phys.~Rev. \textbf{D54}(1996), 1514--1524. (\texttt{hep-th/9602136}).

\bibitem{Ferrara:1996um}
--- \emph{\enquote{{Universality of supersymmetric attractors}}, }Phys.~Rev.
  \textbf{D54}(1996), 1525--1534. (\texttt{hep-th/9603090}).

\end{thebibliography}
%
\end{document}